\documentclass[12pt]{article}

\usepackage{threeparttable}
\usepackage{lscape}

\usepackage{epsfig, verbatim,enumerate, wrapfig}
\usepackage{graphicx}
\usepackage{epstopdf}
\usepackage{subfig}
\usepackage{tikz-cd}

\usepackage{mathrsfs, amsmath}

\usepackage{amsmath,amssymb, amsfonts,euscript,mathrsfs, mathtools}
\usepackage{algorithm,algpseudocode}

\usepackage{pdfpages}

\usepackage{amsthm}

\usepackage{tikz-cd}

\usepackage{biblatex}
\usepackage[compact]{titlesec}

\titlespacing{\paragraph}{0pt}{3pt}{6pt}

\titlespacing{\section}{0pt}{15pt}{*0}
\titlespacing{\subsection}{0pt}{12pt}{*0}
\titlespacing{\subsubsection}{0pt}{12pt}{*0}

\usepackage[left=0.9in, right=0.9in,top=1in, bottom=1in]{geometry}
\usepackage[colorlinks,bookmarksopen,bookmarksnumbered,citecolor=blue,urlcolor=blue]{hyperref} 
\usepackage{float}
\usepackage{multirow}
\floatstyle{plaintop}
\usepackage{amsmath,amssymb, amsfonts,euscript,mathrsfs, mathtools}

\usepackage{bbm}

\usepackage{adjustbox} 

\usepackage{pdfpages}
\usepackage{epstopdf}

\usepackage{pifont}
\usepackage{booktabs}

\allowdisplaybreaks
\begin{document}

\title{Probabilistic Clustering using Shared Latent Variable Model for Assessing  Alzheimer’s Disease Biomarkers}
\author{YIZHEN XU$^1$, SCOTT ZEGER$^2$, ZHEYU WANG$^{2,3, \ast}$, \\[4pt]
\small
1. \textit{Division of Biostatistics, Department of Population Health Sciences, University of Utah, USA}
\\[2pt]
\small
2. \textit{Department of Biostatistics, Johns Hopkins University, USA}
\\[2pt]
\small
3. \textit{Division of Quantitative Sciences, Department of Oncology, Johns Hopkins University, USA}
}

\maketitle
\footnotetext{To whom correspondence should be addressed: wangzy@jhu.edu}

\begin{abstract}
{
The preclinical stage of many neurodegenerative diseases can span decades before symptoms become apparent. Understanding the sequence of preclinical biomarker changes provides a critical opportunity for early diagnosis and effective intervention prior to significant loss of patients' brain functions\cite{hampel2008core}. The main challenge to early detection lies in the absence of direct observation of the disease state and the considerable variability in both biomarkers and disease dynamics among individuals. Recent research hypothesized the existence of subgroups with distinct biomarker patterns due to co-morbidities \cite{butler2021comorbidity} and degrees of brain resilience \cite{arenaza2019metabolic}. Our ability to early diagnose and intervene during the preclinical stage of neurodegenerative diseases will be enhanced by further insights into heterogeneity in the biomarker-disease relationship. 

In this paper, we focus on Alzheimer's disease (AD) and attempt to identify the systematic patterns within the heterogeneous AD biomarker-disease cascade. Specifically, we quantify the disease progression using a dynamic latent variable whose mixture distribution represents patient subgroups. Model estimation uses Hamiltonian Monte Carlo with the number of clusters determined by the Bayesian Information Criterion (BIC). We report simulation studies that investigate the performance of the proposed model in finite sample settings that are similar to our motivating application. We apply the proposed model to the BIOCARD data, a longitudinal study that was conducted over two decades among individuals who were initially cognitively normal. Our application yields evidence consistent with the hypothetical model of biomarker dynamics presented in \cite{jack2013update}. In addition, our analysis identified two subgroups with distinct disease-onset patterns. Finally, we develop a dynamic prediction approach to improve the precision of prognoses. }
\bigskip

\end{abstract}

\section{Introduction}
\label{sec1}

Neurodegenerative diseases cause progressive damage to the brain or peripheral nervous system, impacting crucial functions such as cognition, sensory perception, and motor skills. This process often starts more than a decade before the onset of clinical symptoms; the damage to the brain is considered to be irreversible once symptoms appear\cite{jack2010hypothetical, jack2013update}. Therefore, understanding biomarker dynamics of neurodegenerative diseases before clinical symptoms manifest is critical for developing early diagnostic procedures and facilitating the discovery of more effective interventions \cite{dubois2016preclinical,  vermunt2019duration}. 

A major challenge in modeling biomarker-disease trajectories is the unobservable nature of the pathophysiological process of these diseases and the substantial heterogeneity, both systematic and stochastic, among individuals' biomarker trajectories. Specifically, these diseases often have a long pre-clinical phase where patients are asymptomatic. A definitive diagnosis cannot be made until decades after a pathophysiological process is initiated\cite{small2022early}. Because these disease processes are unobservable, we aim to better understand the biomarker-disease relationship so that we can use biomarker measurements to inform disease processes and make more timely and accurate decisions than the current clinical diagnoses. We must distinguish the pathophysiology of the disease (referred to as the disease) from the clinical diagnosis of the disease (referred to as the diagnosis) in the formulation and modeling of the biomarker-disease relationship. Our application focuses specifically on Alzheimer's disease (AD).  \cite{jack2010hypothetical} hypothesized that the time and shape of biomarker abnormality in AD follow a sequence of sigmoidal curves as the disease progresses through discrete clinical stages. This conceptual model has become widely adopted as the foundation of many AD research. \cite{sperling2011toward} further emphasized that asymptomatic subjects with AD-pathological processes may show subtle signs of impairment in their biomarker measurements. \cite{jack2013update} revised their conceptual framework, replacing the discrete clinical stages with a continuum of ``distance along the pathophysiology pathway''. They highlighted the potential of latent variable modeling using a single latent trait to represent the disease status.

Following this idea, many statistical attempts utilized a latent variable for the underlying disease status when studying AD-related biomarkers relationship. For example, \cite{jack2011evidence} used a cubic spline model with the continuous MMSE score as a surrogate for the disease process. \cite{wang2018biomarker} and \cite{wang2020ad} used unsupervised machine learning to quantify the unobserved disease process as a finite mixture model of continuous biomarkers. \cite{young2014data} studied the sequence of AD biomarker process with a multi-modal change point model. These papers provided some data support in favor of the conceptual framework assumed in \cite{jack2013update}. On the other hand, the results were far from conclusive and revealed a challenge: trajectories of patients' biomarker measurements exhibit substantial heterogeneity. This heterogeneity was emphasized in the National Institute on Aging-Alzheimer's Association workgroups on diagnostic guidelines for Alzheimer's disease \cite{sperling2011toward}. Accordingly, \cite{donohue2014estimating}, \cite{li2019bayesian}, and \cite{sun2019leveraging} assumed biomarkers trajectories to be monotone functions of age with subject-specific time shifts, adjusting for the heterogeneity in disease onset. \cite{jedynak2012computational, jedynak2015computational} formulated a model more closely aligned with the hypothesis \cite{jack2013update} where biomarkers were modeled as sigmoid functions of a latent disease progression score, described by a linear function of patients' age with random intercepts and slopes for individual heterogeneity. 
\cite{tang2023characterizing} further proposed a non-linear mixed effects model that accounts for covariate effects in both the biomarker disease relationship as well as the individual disease process. These studies provided insights into the AD biomarker-disease relationship. However, there remains substantial biomarker heterogeneity that is not explained by the observed covariates or assumed random effects. 

Recent research has unveiled subgroups with systematically different biomarker patterns or progression trajectories which may be attributed to genetic differences, prevalent co-morbidities in the elderly population \cite{butler2021comorbidity}, or the differential brain resilience \cite{arenaza2019metabolic}. For example, \cite{nettiksimmons2010subtypes} used an unsupervised cluster analysis on patients' initial CSF and MRI biomarkers, and identified a subgroup with worse cognitive performance at baseline that progresses faster. On the other hand, GLMMs have been widely used to accommodate substantial heterogeneity in the timing and temporal dynamics of disease progression. For instance, \cite{roy2000latent} proposed a latent time joint mixed effects model for multiple continuous outcomes with a linear Gaussian-distributed latent variable and \cite{proust2006nonlinear} extended the work to allow non-Gaussian biomarkers and specified the latent structure to follow structural equation models. \cite{lai2016multivariate} further extended \cite{proust2006nonlinear} to a mixture model. 

In this paper, we aim to address the systematic and stochastic AD biomarker heterogeneity and discover subpopulations based on their longitudinal disease trajectories. Specifically, we use a continuous latent process to quantify the unobserved latent disease process (LDP) without relying on clinical diagnoses that may be error-prone \cite{small2022early}. We use random effects in the LDP to account for individualized time shifts and temporal correlations. We use a mixture model to account for potential subgroups with varying LDP patterns. We also incorporate patients' demographic and genetic information as regression covariates in both the biomarker model and the LDP to address the biomarker heterogeneity attributed to these factors directly, through the LDP, or both. For example, patients' age is included in the LDP to reflect the significantly higher progressive risk in older patients, as well as in the biomarker model to account for the normal aging effect on patients' biomarker levels aside from the disease. In particular, we adopt generalized linear mixed-effects models (GLMM) \cite{laird1982random} as mixture components in our model. We expand upon \cite{wang2020ad} and \cite{tang2023characterizing}, incorporating GLMMs as mixture components and using clusters to account for systematic variations in the disease risk profiles of subjects. We address the issue of limit-of-detection or boundary effects in the biomarkers by developing a likelihood-based approach. Finally, we create dynamic algorithms that predict patients' most likely future disease trajectory and biomarker value based on their observed history. 

Our paper is organized as follows, section \ref{METHOD} introduces the notations, model, and estimation method. Section \ref{SIM} describes simulation results and demonstrates the validity of the estimation process. Section \ref{APP} focuses on the application of our approach to the BIOCARD (Biomarkers of Cognitive Decline Among Normal Individuals) cohort data and derives BICs to determine the number of clusters. We validate our method with cross-validated trajectory predictions using the computational strategy detailed in the appendices for real-time updating of biomarker trajectories. We also evaluate our method by comparing the rate and duration of diagnostic conversions into MCI and AD conditional on estimated clusters. The findings highlight systematic variability in disease progression across individuals, as well as a temporal ordering of biomarkers that is consistent with the conceptual hypotheses in \cite{jack2013update}. 
 
\section{Method}\label{METHOD}

We adopt the usual terminology of latent variable models. Specifically, the \textit{measurement model} describes the relationship between the observed biomarkers and the latent disease process (LDP). This model addresses the primary scientific focus of this method: the dynamic of biomarker changes along the disease continuum. The \textit{structural model} defines the LDP for each mixture component. In Section 2.1, we follow the hypotheses in \cite{jack2013update}
 and specify the model of biomarker measurements as sigmoid functions of the underlying disease process, which is a mixture representing the heterogeneity in the LDP patterns. We also address the boundary effects of a subset of biomarkers. In Section 2.2, we describe the structural model as the LDP for each mixture component. The method is estimated with a Bayesian approach and the priors are discussed in Section 2.3.

\subsection{The Measurement Model}
Suppose the $i$th subject has $J_i$ time points and $K$ continuous outcomes $\{Y_{ij1},\ldots,Y_{ijK}\}$ at time $t_{ij}, j=1,\ldots,J_i$. We assume $L$ mixture components with different LDP patterns and quantify subject-specific LDP as the continuous latent scores, $d_i(t)=\{d^{(1)}_i(t),\ldots, d^{(L)}_i(t)\}$. For subject $i$, the biomarkers are assumed to be independent conditional on $\{\mu^{(\ell)}_{ik}(t); \ell = 1,\ldots, L, k = 1,\ldots, K\}$, which summarizes the observed characteristics and the true disease status represented by the LDP. The distribution of biomarkers can then be written as
$$f(Y_{i};\mu_i, \sigma) = \sum^L_{\ell=1}\lambda_\ell \bigg[ \prod^{J_i}_{j = 1}\prod^{K}_{k=1}f\bigg(Y_{ijk}; \mu^{(\ell)}_{ik}(t_{ij}),\sigma^{(\ell)}_k\bigg)\bigg],$$
where $\lambda = (\lambda_1,\ldots,\lambda_L)$ is the mixing proportions such that $\lambda_\ell\ge 0$ and $\sum^L_{\ell=1}\lambda_\ell=1$. 

The population-average progression of the $k$th biomarker in the $\ell$th mixture component,  $\mu^{(\ell)}_{ik}(t)$, is modeled as a sigmoid function $h_k(\cdot)$ of the LDP $d^{(\ell)}_i(t)$, along with $X_i(t)^T \beta_k$, which accounts for characteristics-adjusted natural neurodegeneration that occurs besides AD pathology, i.e. $$\mu^{(\ell)}_{ik}(t)=X_i(t)^T \beta_k + h_k(d^{(\ell)}_i(t)).$$ 

We postulate the following pathophysiological pattern of a biomarker $Y_{ijk}$ relative to its latent disease score $d$,
$$h_k(d)= \frac{\gamma_{k1}}{1+ \exp\{-\gamma_{k2} ( d - \gamma_{k3})\}},\qquad \gamma_{k2}>0,$$
assuming that the pattern is shared across mixture components. The sigmoid function $h_k(d)$ depends on the parameters $(\gamma_{k1}, \gamma_{k2}, \gamma_{k3})$; it has an upper bound of $\gamma_{k1}$ when $d\rightarrow \infty$ and a lower bound of 0 when $d\rightarrow -\infty$. Therefore, in an individual without any pathological abnormalities, the average of the $k$th biomarker should be close to the long-term aging effect that is not associated with AD, represented by $X_i(t)^T \beta_k$. However, in individuals who develop pathological abnormalities, we assume that the progression of the average biomarker will follow a sigmoid curve. This curve begins at its initial level in the absence of AD pathological process, $X_i(t)^T \beta_k$, and then progresses towards $X_i(t)^T \beta_k + \gamma_{k1}$. The farthest possible aggravation in the $k$th biomarker due to pathological neurodegeneration is quantified by $\gamma_{k1}$. The parameters $\{\gamma_{k3}; k=1,\ldots, K\}$ are biomarker-specific inflection points in the sigmoid curves and they describe the disease state at which the deterioration in biomarkers is the most apparent. We use inflection points as anchors for studying the temporal order of the biomarkers along the LDP.

Some markers may be censored at boundaries due to the limit of detection or the maximum attainable score in a test. Without loss of generality, we only focus on the ceiling effect. Let $U$ be the subset of biomarkers index where $\{Y_{ijk}; k\in U\}$ are neurological assessments with a ceiling effect, e.g. we observe value $u_k$ when in fact $Y_{ijk} \ge u_k$. For example, most cognitive tests have a maximum score, imposing a limit to how much cognitive capability one can measure. This issue is particularly prevalent when studying patients who are in the pre-clinical or early stages of disease progression, as their cognitive functions are still largely intact. This leads to a skewed distribution of measurements, with a concentration of high scores that deviates from a normal distribution. We assume that the biomarkers and test scores follow normal distributions, and model the ceiling effect as follows,
\begin{align*}
   & f\bigg(Y_{ijk}; \mu^{(\ell)}_{ik}(t_{ij}),\sigma^{(\ell)}_k\bigg) \\
=&\begin{cases}
 \phi(\frac{Y_{ijk}- \mu^{(\ell)}_{ik}(t_{ij})}{\sigma^{(\ell)}_k}) & \text{ if } k\in U^c \\ 
\mathbbm{1}\{Y_{ijk}=u_k\}\int_{y_{ijk}\ge u_k} \phi(\frac{y_{ijk}- \mu^{(\ell)}_{ik}(t_{ij})}{\sigma^{(\ell)}_k})d y_{ijk} + \mathbbm{1}\{Y_{ijk}<u_k\}  \phi(\frac{Y_{ijk}- \mu^{(\ell)}_{ik}(t_{ij})}{\sigma^{(\ell)}_k}) & \text{ if } k\in U
\end{cases},
\end{align*}
where $\phi(\cdot)$ is the standard normal density function.


\subsection{The Latent Structural Model}
Suppose that there are $L$ distinct LDP patterns in the population. We then specify the LDP in the $\ell$th mixture component at time $t$ for individual $i$, which is shared among the $K$ biomarkers, as follows, 
$$d^{(\ell)}_i(t) = -\alpha^{(\ell)}_0 + Z_i(t)^T \alpha + \theta^{(\ell)}_i(t), \qquad t\ge 0,i\in\{1,\ldots,N\},$$
where $Z_i(t)$ is a $q$-dimentional vector of the observed factors that are known to be associated with the underlying AD pathophysiology, such as patients' age and the number of APOE $\epsilon$-4 allele. In particular, $Z_i(t)$ may overlap with the set of covariates in $X_i(t)$. 

The intercept $\alpha^{(\ell)}_0$ represents the unique offset in the LDP for the $\ell$th cluster. It captures differences in systematic subject profiles caused by unobserved factors in the progression of AD, such as comorbidities or brain resilience. A higher value of $\alpha^{(\ell)}_0$ results in a decreased disease score. It can be viewed as a resilience against AD in the risk profile. Similar to \cite{sun2019leveraging}, we assume the long-term profile parameter to be constant for a relatively short observation time window. For model identifiability and to prevent label switching, we assume $\alpha^{(1)}_0=0 < \ldots < \alpha^{(L)}_0$, where a larger cluster index indicates a lower risk profile.

To account for subject-level heterogeneity in the temporal dynamics of the LDP, we assume that the time-varying random effects $\theta^{(\ell)}_i(t)$ are Gaussian processes so that potential nonlinearity over time can be captured by kernel functions.
The process $\theta^{(\ell)}_i(t)$ concentrates round zero with variation quantified by the exponentiated quadratic covariance function, 
$$k(\theta^{(\ell)}_i(t), \theta^{(\ell)}_i(t'); \tau, \rho^{(\ell)}) =\tau^2\exp\bigg\{-\frac{1}{2\rho^{(\ell)2}} (t-t')^2\bigg\}, $$
where marginal deviation $\tau$ is set to one for identifiability, e.g. $\text{var} (\theta^{(\ell)}_i(t)) = 1$ for any $t\ge 0$, and covariance $k(\cdot, \cdot;1,\rho^{(\ell)} ) $ equals correlation. Here, the random effects of a subject at any two time points have a stationary positive correlation, which is completely determined by the time difference. A larger length scale, $\rho^{(\ell)}$,  indicates a faster decay in the correlations.

\subsection{Prior Specification}
Parameters in the measurement model are independently assigned weakly informative proper priors, $\beta_k \sim N(0,100)$ and $\sigma_k \sim \text{Inv-gamma}(0.01, 0.01)$. Choosing priors for the structural model parameters requires more consideration. Considering that the age variable is standardized in our analysis and that $\theta^{(\ell)}_i(t) \sim N(0,1)$, we know that the random effects $\theta^{(\ell)}_i(t)$ and time, i.e. the transformed age, are mostly in  $(-2,2)$. Note that any time difference is between $(-2)-2=-4$ and $2-(-2)=4$. Write the coefficient for time in the LDP as $\alpha_{t}$, we can then view $\theta^{(\ell)}_i(t) / \alpha_{t}$ as a subject-record-specific time shift in disease progression,  $$\alpha_{t} t + \theta^{(\ell)}_i(t) = \alpha_{t}[t+\theta^{(\ell)}_i(t) / \alpha_{t}].$$ Thus, the time shift is mostly in $(-2/\alpha_{t},2/\alpha_{t})$ and this interval should mostly include all possible values of time differences, such as $(-4,4)$. A reasonable proper prior for $\alpha_{t}$ is $N(0,1)$ because it implies that the parameter largely satisfies $|\alpha_{t}| \le 2$ and hence $(-4,4)\subseteq (-2/\alpha_{t},2/\alpha_{t})$ with a high probability. Similar reasoning applies to the coefficient of other covariates in the LDP. 

Following the discussion in  \cite{tang2023characterizing}, we acknowledge that the pathophysiology of AD spans several decades and that the BIOCARD study in our application does not encompass the entire disease development process for all subjects. Therefore, we pre-determine the maximum extent of possible AD-related biomarker changes, $\gamma_{11},\ldots,\gamma_{K1}$. This specification is based on the assumption that early-progressing markers, such as CSF and MRI, have likely shown the maximum possible extent of progression during the study period and that late-progressing markers, such as cognitive tests, are standardized across studies and are relatively well understood. To suppress degeneration of the sigmoid curves in a finite study sample, $\gamma_{12},\ldots,\gamma_{K2}$ are given weakly informative gamma prior with shape parameter 3 and rate 1. For inflection points $\gamma_3 = \{\gamma_{13},\ldots,\gamma_{K3}\}$, we assume hierarchical priors conditional on the type of measure the $k$th biomarker belongs to. For instance, MRI-related markers simultaneously reflect brain structure and are considered to have similar AD pathophysiology patterns relative to disease progression. Hence, $\gamma_3 = \{\gamma^{type}_3; type \in \{COG, MRI, CSF\}\}$ has priors $\gamma^{type}_3 \sim N(\mu^{type},1)$ and $\mu^{type}\sim N(0,2^2)$. This hierarchical prior specification allows for variability in the sampling of $\gamma_3$ because the prior distribution spans the range $(-6,6)$ with high probability, which overlaps with the support of the latent scores sufficiently well. This is important because an inflection point anchors the location on an LDP where a biomarker deteriorates the fastest. Since inflection points are important anchors within the range of latent scores, we further restrict the shift of latent scores across clusters to remain inside the scope of $\mu^{type}$ by specifying the prior distribution for the risk profile parameter $\alpha^{(\ell)}_0$, $\ell > 1$, to be a gamma distribution with shape 2 and rate 1.5.

In the latent structural model, when $t$ and $t'$ are $\rho^{(\ell)}$ apart, the correlation between $\theta^{(\ell)}_i(t)$ and $\theta^{(\ell)}_i(t')$ is  $\exp(-\frac{1}{2})\approx 0.61$. Define $\Delta_i$ to be the maximum time difference among observed time points of the $i$th subject, e.g. $\Delta_{i} = \underset{1 \leq j \leq J_i}{\max}(t_{ij})-\underset{1 \leq j \leq J_i}{\min}(t_{ij})$. We suppress the length scales $\rho^{(\ell)}$ to be no larger than the maximum distance in observed time, $\underset{1 \leq i \leq N}{\max} (\Delta_i)$, so that a correlation higher than 0.61 is unlikely to occur between two time points that are farther apart than $\underset{1 \leq i \leq N}{\max} (\Delta_i)$. Similarly, the length scale is no smaller than the minimum distance in observed time, $\underset{1 \leq i \leq N}{\min} (\Delta_i)$. To achieve this, we assume a prior $\rho^{(\ell)} \sim \text{inv-gamma}(a_\rho, b_\rho)$ such that $P(\rho^{(\ell)}  \in [\min(\Delta_t),\max(\Delta_t)]) =98\%$, where hyper-parameters $a_\rho$ and $b_\rho$ are calculated based on the data.  

\section{Applications - BIOCARD Cohort}\label{APP}

 
 The motivating application is to investigate the Biomarkers of Cognitive Decline Among Normal Individuals (BIOCARD) study. This is a longitudinal observational study that aimed to identify early markers of cognitive decline and AD progression in cognitively normal individuals. Individuals were initially cognitively normal with an average enrollment age of around 50. More than half of the participants had a first-degree relative with dementia. The observational study records biennial measurements of magnetic resonance imaging (MRI) scans and cerebrospinal fluid (CSF), as well as annual clinical assessments and cognitive tests. The assessments classify individuals as CN, MCI, or dementia. Further study details can be found in \cite{albert2014cognitive} and at \url{https://www.biocard-se.org}. 
 
At the time of recruitment, all participants were between the ages of 20 and 86 and had a minimum of twelve years of education. Patients' demographic and medical information was collected at the baseline. Table \ref{table:baseline} contains the following information: (1) a description of the 11 biomarkers being studied jointly, with summary statistics across all patients and stratified by observed conversions of AD diagnoses, i.e. progression from CN to cognitively unaltered, MCI, or AD at the end of follow-up; (2) a summary of the baseline risk variables; and (3) the average observed conversion time for those who were assessed to have MCI or AD before data closure.  Cognitive test scores, entorhinal cortex thickness, and other MRI/CSF measures of interest had approximately less than 5\%, 12\%, and 20\% missingness, respectively. After the initial visit, most of the follow-up visits happened every 12 months. The measurement model includes risk covariates age, gender, presence of ApoE 4 alleles, and education. The risk factors in the latent structure model are age, ApoE 4 alleles, and the interaction between them.

\begin{figure}[!htb]
\centering
        \includegraphics[scale = 1]{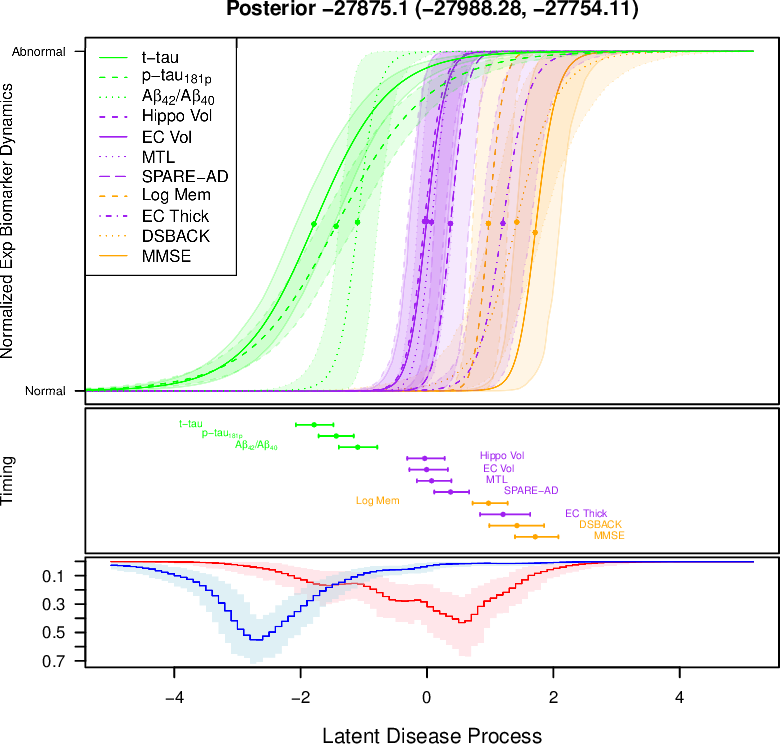}
        \caption{Under $L=2$, the top figure is the estimated $\exp(h_k(d))$ as a function of latent score $d$ for the 11 biomarkers, normalized to the range of [0,1], where dots represent the inflection points of the exponentiated growth curves, solid lines represent the posterior mean estimates and bands reflect the 95\% credible intervals, and green, purple, and orange correspond to CSF, MRI, and cognitive measurements, respectively; the middle figure is the summary of corresponding inflection points, showing the biomarkers' temporal sequence and its relative uncertainty on the exponential scale; the bottom figure displays the estimated distribution of $\{d_i^{(\ell)}(t_{ij});  j=1,\ldots,J_i,\text{ and } i=1,\ldots,N\}$, where blue $(\ell = 1)$ is the subgroup that was less progressed in AD and red $(\ell = 2)$ is the subgroup with higher risk profile.}
\label{fig:curvesL2}
\end{figure}

\begin{figure}[!htb]
\centering
        \includegraphics[scale = 0.9]{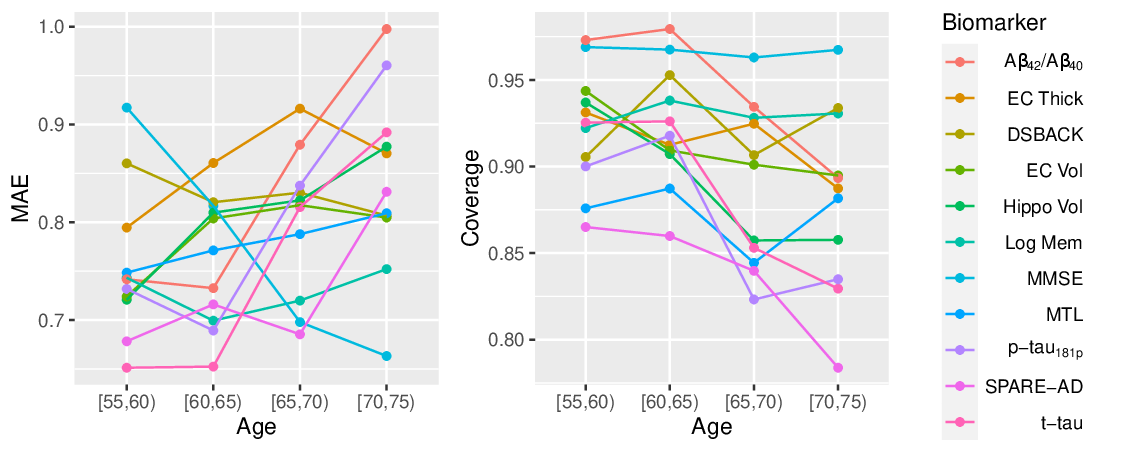}
        \caption{Under $L=2$, the mean absolute error (MAE) and posterior predictive coverage of the out-of-sample prediction of all individuals averaged across 5-fold cross-validation. The x-axis displays age intervals, the range of which approximately covers the 15\% to 85\% quantiles (55.6 to 76.6 years) of the observed age in the BIOCARD data.}
\label{fig:accuracyL2}
\end{figure}

We use the following Bayesian Information Criterion (BIC) approximation \cite{delattre2014note,shen2021bayesian} for model selection,
$$BIC = -2 \log f(\hat{\psi}|\mathbf{Y})+\dim(\psi) \log (N),$$
where $\psi = \{\lambda_\ell, \beta_k, \gamma_k, \sigma^{(\ell)}_k, \rho^{(\ell)}, \alpha, \alpha^{(\ell)}_0; \ell = 1,\ldots, L, \text{ and }k=1,\ldots,K\}$, $\hat{\psi}$ is the maximum a posteriori (MAP) estimate of the parameters, and $\dim(\cdot)$ is the dimension of a vector. Because the prior densities are not uniform, the posterior probability $f(\hat{\psi}|\mathbf{Y})$ is used instead of the likelihood $\mathcal{L}(\hat{\psi}|\mathbf{Y})$ at MAP in the formula. The BIC is calculated to be 57.4, 55.9, and 58.2 ($\times 10^3$) for $L = $1, 2, and 3. Since a lower BIC suggests a better model, we chose probabilistic clustering under $L=2$ as the final strategy.  Under $L=2$, the estimated biomarker-pathophysiological patterns are visualized in Figure \ref{fig:curvesL2}, and the cross-validated predictive accuracies over different age categories are displayed in Figure \ref{fig:accuracyL2}. Under $L=2$, the estimated parameters in the measurement model, the latent structural model, and mixing proportions are summarized in Part 3 of the Supplementary Materials.

We examine the estimated biomarker-pathophysiological patterns of the 11 measures of interest, as shown in Figure \ref{fig:curvesL2}. The estimation results under $L=2$ are consistent with relevant scientific evidence on the temporal ordering of biomarkers described in \cite{jack2013update}: first, A$\beta_{42}$/A$\beta_{40}$ and t-tau are the earliest major AD biomarkers to become aberrant, followed by structural MRI and clinical symptoms; second, A$\beta_{42}$/A$\beta_{40}$ was completely abnormal years before diagnosis of dementia;
third, p-tau$_{181p}$ and t-tau progresses similarly over time, becoming increasingly abnormal as the disease progresses;
and fourth, A$\beta_{42}$/A$\beta_{40}$ is more abnormal early in the course of the disease than hippocampus volume. In addition, the resilience towards AD of individuals in group 2 in terms of time can be estimated by $\alpha^{(2)}_0/\alpha^{(2)}_t$. This quantity has a posterior mean of 5.6 years with a 95\% credible interval (4.03, 7.92).

We investigate the dynamic patterns of AD clinical diagnoses, especially subjects' status conversion into MCI or AD, stratified by the risk profiles estimated from the proposed L-group probabilistic clustering. The diagnoses are not used as labels for supervised learning as they might be imprecise and unstable during transition periods. However, diagnostic-based time of state conversion is of essential scientific relevance, so we utilize it as a metric for validating and comparing cluster findings. Let a subject's initial diagnosis of a disease state be the first available clinical assessment of that state. We define the conversion time from CN to a non-CN final state to be the time between the baseline and when a matching diagnostic first occurred during a subject's follow-up.

Under the selected probabilistic clustering model ($L=2$), Table \ref{table:1stconversion} provides posterior cluster membership-stratified summaries of subjects' patterns of developing a more severe disease state, e.g., going from symptom-free to MCI and from MCI to AD. Based on an individual's complete trajectories, the posterior cluster membership is determined to be the cluster index with the highest marginal posterior probability. We observe that subjects in the second group, that is, the cluster assessed to have a systematically lower risk profile, have a smaller ratio of developing MCI or AD, significantly fewer cases of AD than MCI, and a longer duration to acquire cognitive impairment. Results show that subjects in the lower-risk profile group generally have a lower rate and longer duration of disease aggravation.

In addition to the conversion time between diagnostic statuses, we evaluate the approach using the accuracy of the out-of-sample trajectory predictions. To start with, we create five folds among all individuals. Then, for the $k$th fold, we build a $L=2$ probabilistic clustering model based on the other four folds and obtain estimates for parameters $(\beta, \alpha_0, \alpha,\sigma,\rho)$, denoted as $\hat{\psi}_k$. Next, we make time-varying predictions of the trajectories as follows. For individual $i$ at the $j$ time point, we condition on $Y_{i,1:j}$ to predict $Y_{i, j+1}$, i.e. obtain the posterior predictive distribution of the outcome at time $t_{i,j+1}$ using the person's history trajectories. We first estimate the posterior predictive distribution of cluster probability, $P(\text{Cluster }\ell| Y_{i,1:j}; \hat{\psi}_k)$ for $\ell = 1,2$, following the numerical approximation approach outlined in Appendix 1. Next, based on the computational strategy proposed in Appendix 3, we can represent the posterior predictive distribution of $P(Y_{i,j+1} |Y_{i,1:j};\hat{\psi}_k)$ by $\{\hat{Y}^{(r)}_{i,j+1}; r = 1,\ldots, N_{post}\}$. Based on the posterior mean and 95$\%$ credible interval of the predictive samples, we aggregate across all individuals and summarize the mean absolute error (MAE) and the posterior coverage over different age categories in Figure \ref{fig:accuracyL2}. The MAEs are less than one standard deviation from the observed values for all eleven biomarker observations that fell roughly between the 15\% and 85\% quantiles of the overall age distribution, or between 55 and 75 years old, and the posterior coverages are primarily greater than 80\%.

\section{Discussion}
In summary, our work uses a continuous latent metric to capture the biological AD continuum based on information shared by multiple biomarkers that are repeatedly measured. Since there is no gold standard for AD neuropathologic status during the preclinical stages, the latent metric could be used as a composite score to guide early AD detection. Our work also extends the latent structure specification in \cite{jedynak2012computational}  to incorporate covariates' impact on the pathological development, as well as Gaussian process-distributed random effects to account for potential subject-specific time-varying heterogeneity and autocorrelation. Furthermore, this paper provides a unified framework for simultaneously identifying subgroups by systematic disease risk profiles, and determining the temporal order of biomarker changes along a latent metric while accounting for boundary effects. Such a framework allows for a more individualized examination of the biomarker-disease relationship and the identification of distinct group patterns. This method has the potential to be utilized in clinical trials to identify and recruit asymptomatic individuals who are more likely to develop AD symptoms in the near future. Finally, as illustrated in our application, we present a computational strategy for making real-time updates to trajectory predictions using the proposed GLMM mixture model. With Bayesian estimation, our method quantifies the uncertainty in determining the temporal sequence of biomarkers and making dynamic trajectory predictions.

Using the BIOCARD data, this study proposed a novel unsupervised machine learning approach for (1) generating an AD risk score among initially CN individuals to describe the progression of AD, (2) identifying different groups of subjects with systematically distinct risk profiles of developing AD symptoms, and (3) discovering the temporal sequence of multiple biomarkers as AD progresses. In terms of external validation, clusters identified based on latent risk profiles well separate diagnostic-based status conversions into MCI or AD. The results show that the AD risk score, the LDP from the unsupervised machine learning, captures the underlying disease burden well and can potentially be applied to individuals in other stages of the AD continuum.

The proposed method has several advantages. The probabilistic clustering approach allows for the early detection of asymptomatic subjects who have a systematically higher risk profile of developing AD; this has the potential to significantly benefit AD-related clinical trial designs, allowing studies to more efficiently recruit asymptomatic subjects who have a higher chance of developing AD sooner. A byproduct of the approach is the identification of the chronological sequence in which biomarkers exhibit the most noticeable changes in relation to the progression of AD. The findings align with existing hypotheses and research. Furthermore, the latent variable framework provides greater flexibility and capability in accounting for more aspects of the disease mechanism, allowing for a more robust description of the underlying pathophysiology. The method can also be easily modified to include new biomarkers such as medical co-morbidities and genetic sequencing. The Bayesian approach offers additional benefits by naturally quantifying the stochastic uncertainty in the estimation of cluster membership, the temporal order of biomarkers by AD progression, and the dynamic prediction of trajectories.

One limitation of this study is that the clustering divides the population into groups with relatively homogeneous AD progression patterns. Our method does not identify clusters based on a gold standard disease outcome or on AD clinical diagnoses, which are imprecise surrogates for true AD status. The estimated LDP is considered to reflect AD pathophysiology because biomarkers are selected to be biologically relevant and, as with AD pathology, the LDP is the essential shared factor that contributes to the heterogeneity of biomarkers. Furthermore, the method is limited to subjects who were initially asymptomatic; the LDP lacks an anchor that allows application on data from subjects who were cognitively abnormal at enrollment.

Possible extensions of the proposal include (1) use semi-parametric transformation of the outcomes with link functions to allow for non-Gaussian biomarkers of various types; (2) specify a threshold parameter in the LDP as an anchor on clinical diagnoses of cognitive abnormality; (3) jointly model the longitudinal biomarkers with competing risks events such as depression and death; and (4) allow covariates to also contribute to mixing proportions and relax the assumption of linear trends in covariates.

\section*{Acknowledgment}
 This work is supported by the National Institutes of Health grant R01AG068002. The content is solely the responsibility of the authors and does not necessarily represent the official views of the National Institutes of Health.

\newpage

\bigskip

\baselineskip 22pt

\printbibliography

\newpage

\section*{Tables}

\begin{table}[htp]
\begin{tabular}{ccccc}
\hline\hline
                & All (N=297)         & NM-NM (N=215)        & NM-MCI (N=53)        & NM-AD (N=29)        \\ \cline{2-5} 
Conversion Time &                     &                      & 12.61 (4.25, 23.12)  & 13.45 (3.01, 19.17) \\
Age             & 57.21 (31.6, 76.76) & 55.45 (28.87, 73.44) & 60.06 (36.99, 73.61) & 65.1 (38.55, 80.18) \\
Apoe            & 0.61                & 0.60                  & 0.57                 & 0.69                \\
Gender          & 0.59                & 0.60                  & 0.62                 & 0.41                \\
Education       & -0.07 (-2.23, 1.2)  & -0.02 (-2.23, 1.2)   & -0.3 (-2.23, 1.2)    & -0.06 (-2.23, 1.2)  \\
MMSE Score     & -0.27 (-0.59, 1.54) & -0.3 (-0.59, 0.83)   & -0.17 (-0.59, 1.54)  & -0.18 (-0.59, 0.83) \\
Log Mem          & 0.69 (-1.12, 2.77)  & 0.57 (-1.12, 2.53)   & 0.99 (-0.63, 2.77)   & 1.02 (-1.12, 2.53)  \\
DSBACK            & 0.14 (-1.7, 1.86)   & -0.04 (-1.86, 1.86)  & 0.56 (-0.91, 1.94)   & 0.62 (-0.43, 1.55)  \\
EC Thick       & -0.02 (-1.58, 2.04) & -0.11 (-1.74, 1.74)  & 0.09 (-1.49, 2.04)   & 0.48 (-1.11, 1.96)  \\
Hippo Vol  & -0.16 (-1.77, 1.66) & -0.22 (-1.84, 1.43)  & -0.16 (-1.67, 1.56)  & 0.28 (-1.45, 1.49)  \\
EC Vol    & -0.04 (-2.07, 1.56) & -0.04 (-1.44, 1.45)  & -0.08 (-2.77, 1.48)  & 0.06 (-1.74, 1.69)  \\
MTL            & -0.16 (-2.08, 1.79) & -0.19 (-1.93, 1.43)  & -0.17 (-2.24, 1.84)  & 0.11 (-2.12, 2.31)  \\
SPARE-AD       & -0.25 (-1.8, 1.47)  & -0.31 (-1.79, 1.28)  & -0.18 (-1.8, 0.95)   & 0.09 (-1.3, 1.57)   \\
T-tau            & -0.23 (-1.18, 2.17) & -0.37 (-1.25, 0.78)  & -0.05 (-0.95, 1.74)  & 0.62 (-1, 2.69)     \\
P-tau$_{181p}$         & -0.24 (-1.16, 2.27) & -0.39 (-1.17, 0.98)  & -0.06 (-1.05, 2.24)  & 0.6 (-0.89, 2.67)   \\
A$\beta_{42}$/A$\beta_{40}$        & -0.12 (-1.23, 1.79) & -0.28 (-1.33, 1.64)  & 0.21 (-0.89, 1.68)   & 0.56 (-0.92, 1.74)  \\ \hline\hline
\end{tabular}
\caption{Summary of covariates and biomarkers at the baseline for all subjects and three subgroups of those who stayed normal, obtained MCI, and AD at the time of data closure. We jointly studied the 11 biomarkers: CSF biomarkers include beta-amyloid (A$\beta$) 42 over 40 ratio (A$\beta_{42}$/A$\beta_{40}$), p-tau$_{181p}$, and total tau (t-tau);  MRI biomarkers include entorhinal cortex thickness (ECT), entorhinal cortex volume (ECV), hippocampal volume (HV),  medial temporal lobe (MTL) composite score \cite{pettigrew2017progressive}, and SPARE-AD score (Spatial Pattern of Abnormalities for Recognition of Early AD)\cite{davatzikos2009longitudinal}; cognitive scores include Digit Span Backward (DSBACK) from the Wechsler Adult Intelligence Scale-Revised \cite{wechsler1997psychological, wang2020ad}, Logical Memory (LM) delayed recall from the Wechsler Memory Scale-Revised, and MMSE (Mini-Mental State Examination) score. MRI biomarkers are computed as an average over the left and right hemispheres. In addition, the volumetric measures are adjusted for intracranial cavity volume by division. 
}
\label{table:baseline}
\end{table}

\begin{table}[htp]
\centering
\begin{tabular}{cccc}
\hline\hline 
Cluster Index & Conversion Type & Percentage        & Duration            \\ \hline
1             & CN-MCI          & 0.40 (0.38, 0.42) & 10.40 (10.18, 10.67) \\
             & CN-MCI'          & 0.40 (0.38, 0.42) & 10.38 (10.18, 10.60)  \\
              & MCI-AD          & 0.44 (0.41, 0.46) & 3.25 (3.18, 3.36)    \\
2             & CN-MCI          & 0.26 (0.25, 0.27) & 12.17 (11.99, 12.28) \\
             & CN-MCI'          &  0.20 (0.19, 0.20) &  13.36 (13.14, 13.52)  \\
              & MCI-AD          & 0.15 (0.13, 0.17) & 3.81 (3.55, 3.92)    \\ \hline\hline
\end{tabular}
\caption{Summary of conversion from the first diagnosis of a state A to the first diagnosis of a state B for subjects in each cluster from the probabilistic clustering model under $L=2$, where states $(\text{A, B})\in \{(\text{CN, MCI}),(\text{CN, MCI'}), (\text{MCI, AD})\}$. Note that MCI' represents confirmed MCI status where an individual's last diagnosis is not normal, e.g. a person with three consecutive diagnoses of CN-MCI-CN does not fall into this category. The posterior mean and 95\% credible interval of the percentage and duration of conversion are calculated. Percentage is defined as the ratio of subjects who developed state B among those who were in state A.}
\label{table:1stconversion}
\end{table}

\section*{Figures}

 \end{document}